\documentclass[twocolumn,showpacs,preprintnumbers,amsmath,amssymb]{revtex4}

\usepackage{graphicx}
\pdfoutput=1

\begin{document}


\title{Photoluminescence clamping with few excitons in a single-walled carbon nanotube}

\author{Y.-F. Xiao, T.Q. Nhan, M.W.B. Wilson, and James M. Fraser}
\affiliation{Department of Physics, Engineering Physics \& Astronomy, Queen's University, Kingston, Ontario, K7L 3N6 Canada }%
\date{\today}

\begin{abstract}

Single air-suspended carbon nanotubes (length 2 - 5 $\mu$m)
exhibit high optical quantum efficiency (7 - 20\%) for resonant
pumping at low intensities.  Under ultrafast excitation, the
photoluminescence dramatically saturates for very low injected
exciton numbers (2 to 6 excitons per pulse per SWCNT).  This PL
clamping is attributed to highly efficient exciton-exciton
annihilation over micron length scales.  Stochastic modeling of
exciton dynamics and femtosecond excitation correlation
spectroscopy allow determination of nanotube absorption (2 - 6\%)
and exciton lifetime (85 $\pm$ 20 ps).
\end{abstract}

\maketitle

Understanding the electronic and optical properties of
single-walled carbon nanotubes (SWCNTs) will no doubt benefit the
development of carbon-based optoelectronic
devices~\cite{Misewich_Sci03,Freitag_NL03,Star_NL04}. Great
advances have been made in synthesizing SWCNTs of higher quality
and isolating them from perturbations~\cite{OConnell_Sci02,
Hertel_PSS06,Islam_PRL04} with a general trend from measurement on
ensemble towards single tube
level~\cite{Lefebvre_PRB04,Hagen_PRL05,Hogele_PRL08,Weisman_Sci07}.
For encapsulated SWCNTs, exciton many-body interactions manifest
themselves in sublinear photoluminescence emission (PL) and is
well explained by diffusion limited exciton-exciton
annihilation~\cite{Russo_PRB06,Murakami_PRL09}, with diffusion
lengths of 6 - 90 nm~\cite{Luer_NP09,Weisman_Sci07}.  The recent
observation of photon antibunching in SWCNT PL is also consistent
with exciton localization or very efficient exciton-exciton
annihilation~\cite{Hogele_PRL08}.  It has been shown that
processing affects linear properties~\cite{Berciaud_PRL08}, but
the effect that sonification and encapsulation might have on
exciton interactions is not well understood.

By studying a single air-suspended, unprocessed
SWCNT~\cite{Lefebvre_PRB04,Lefebvre_NL06}, we hope to reduce
ambiguities posed by environmental effects and isolate the
inherent SWCNT properties and exciton dynamics. In this letter, we
report on studies of PL fluence (emission from E$_{11}$ van Hove
singularity excitons) as a function of pump fluence (resonant to
E$_{22}$ excitons) on single (9,8) and (10,8) SWCNTs, using
optical 150 fs pulse, 4 ps pulse, and continuous-wave (CW)
excitation. Unperturbed SWCNTs are studied by selecting relatively
bright SWCNT that exhibit narrow room temperature emission ($\sim
12$ meV) and absorption ($\sim 44$ meV) linewidths.  They are also
sufficiently long that we can resolve their lengths using high
resolution PL mapping.  Under low excitation, PL output relative
to incident fluence shows a linear dependence and a large PL
action cross section (defined in Ref.~\cite{Tsyboulski_NL07})
consistent with high quality SWCNTs. Also similar to encapsulated
SWCNTs, exciton linear decay lifetimes are 85 $\pm$ 20 ps.  In
sharp contrast to encapsulated SWCNT behavior, the PL from an
air-suspended SWCNT saturates at moderate pump fluence (100
photon/pulse/SWCNT) and does not increase for over an order of
magnitude pump increase.  Comparison to a stochastic model of
exciton decay shows PL clamping is consistent with efficient
exciton-exciton annihilation over micron-length scales.  This is
considerably longer than expected based on models developed for
SWCNTs (exciton diffusion lengths 6 - 90
nm~\cite{Luer_NP09,Weisman_Sci07}), suggesting that exciton
interactions are inherently different in an air-suspended SWCNT.

Air-suspended SWCNTs are grown by chemical vapor deposition on
lithographically patterned silicon dioxide on silicon wafers
(similar to Ref.~\cite{Kaminska_NT07}). The SWCNTs are suspended
on 0.5 $\mu$m high ridges. During experimentation, the SWCNT
environment is purged with dry N$_2$ to avoid sample degradation.
PL spatial mapping is used to locate bright (9,8) and (10,8)
SWCNTs and indicates low SWCNT density($\sim$ 3 bright SWCNTs per
mm$^2$). Photoluminescence excitation spectroscopy and high
resolution PL mapping are performed on candidate SWCNT to
determine species, quality, length and orientation. SWCNTs
analyzed in this letter maintained their PL emission as well as
spectral shape throughout the entire investigation.
\begin{figure*}
\centering
\includegraphics[width=1\textwidth]{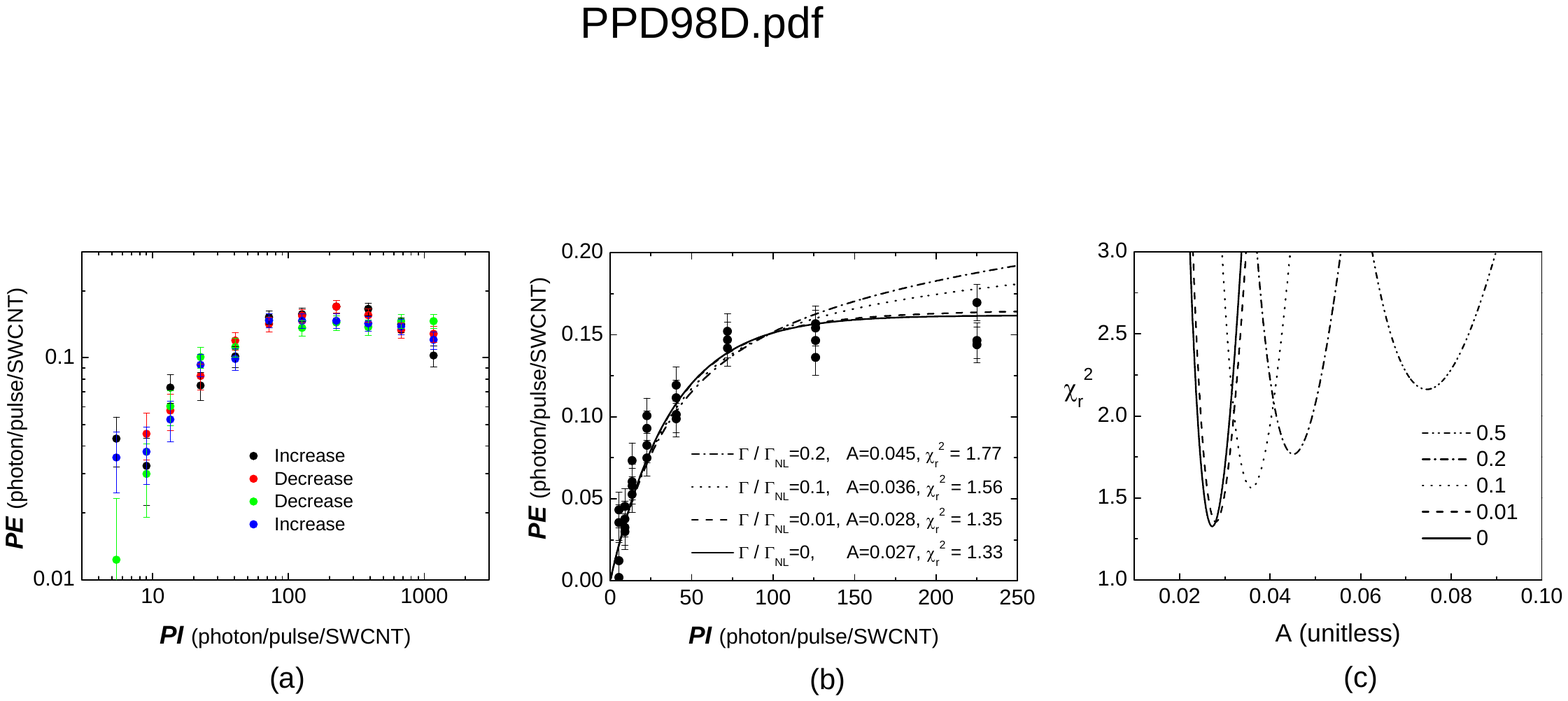}
\caption{(a) Single (9,8) SWCNT PL as a function of incident pump
fluence (recorded using both increasing and decreasing pump
fluences). (b) Experimental data (dots) are compared to results
from a stochastic model (solid lines) with various
$\frac{\Gamma}{\Gamma_{NL}}$ ratios and absorption coefficients
(A). (c) Reduced chi-square as a function of A are calculated at
different ratios of $\frac{\Gamma}{\Gamma_{NL}}$. }
\label{fig:PPD98D}
\end{figure*}

Results from a typical (9,8) SWCNT (length 4.5 $\mu$m) are shown
in Figure~\ref{fig:PPD98D} (a). Optical excitation is by 150 fs
duration pulses (Ti:sapphire oscillator at 800 nm center
wavelength, 76 MHz repetition rate, linear polarized parallel to
SWCNT axis).  The SWCNT is quasi-homogeneously excited by setting
the excitation beam FWHM ($13$ $\mu$m) larger than the tube
length. Total number of photons incident, $PI$, is determined from
the product of pump intensity and surface area of the unrolled
tube. Multiple scans for alternatively increasing and decreasing
pump fluence show no hysteresis or aging.  Photons emitted, $PE$,
is the total photon number emitted per pulse per
SWCNT~\cite{calibration}.  PL in the region around saturation
onset is shown in Figure~\ref{fig:PPD98D} (b) on a linear scale.
At low pump fluence, the slope of $PE$ relative to $PI$ matches CW
excitation and its value is the PL action cross section (0.004)
which is within a factor of three (higher) compared to previous
high quality encapsulated samples~\cite{Tsyboulski_NL07}.  With
higher pump fluence, the PL deviates from linear and reaches an
onset point ($PI \sim 100$ or $7\times 10^{11}$
photon/cm$^2$/pulse) where $PE$ is fully saturated.  Most past
work on nanotube ensembles reported onset of sublinear behavior at
much higher pump fluences (2 - 3 orders of magnitude) and did not
observe PL clamping~\cite{Hagen_PRL05,Ma_MP06,Murakami_PRL09}.
Nonlinear behavior was attributed to exciton-exciton interactions
resulting in nonradiative annihilation.  A similar hard saturation
behavior was recently reported for SDBS-encapsulated SWCNTs
measured in confocal configuration by H\"{o}gele \textit{et
al.}~\cite{Hogele_PRL08} at much higher pump fluence (10$^{14}$
photon/cm$^2$/pulse) but they employed a pump resonant to a phonon
sideband.  In our studies, we have noted that the strong PL
clamping is lost when we employ inhomogeneous excitation (smaller
pump spot than tube length)~\cite{Xiao_SPIE09}.  A similar problem
would occur for ensemble studies unless the field of view was
restricted to the region of the sample that is quasi-uniformly
excited (\textit{e.g.} as provided by confocal light collection).
An additional problem for random orientated ensemble systems is
that the light field couples with variable efficiency depending on
SWCNT orientation thus lessening saturation effects for similar
intensities.  Even so, inhomogeneous excitation or coupling is not
expected to account for the dramatic differences in nonlinear PL
behavior between air-suspended and encapsulated SWCNTs.  A careful
examination of the air-suspended SWCNT behavior is required to
ascertain if the microscopic processes affecting exciton dynamics
are fundamentally different compared to encapsulated samples.

A first question is if air-suspended SWCNTs demonstrate different
linear behavior for absorption or emission.  PL action cross
section is a useful metric since measurement of absorption
independently of emission efficiency is not
straightforward~\cite{Tsyboulski_NL07}.  By attributing the
observed saturation to exciton-exciton annihilation, we can
separate the relative contributions to PL action cross section
($A\times \alpha\eta_{QE}$) of absorption (unitless $A$) and
emission (experimentally resolved $\alpha\eta_{QE}$).  $A$ is the
product of the atomic absorption cross section and carbon surface
density~\cite{Lefebvre_NL06}. $\eta_{QE}$ is the intrinsic optical
quantum efficiency and $\alpha \leq 1$ due to nanotube
imperfections.  The results from a detailed stochastic model are
described below but it is instructive to determine an upper limit
for absorption by assuming annihilation is instantaneous over the
entire SWCNT length.  Thus one or more excitons relaxes quickly to
just one exciton, which then decays through linear recombination
including radiative emission.  Under this simple model, since the
injection of one or more excitons produces identical PL, the full
time dynamics do not need to be modeled to find PL power
dependence; all that is relevant is the probability of injecting
zero excitons per pump pulse~\cite{Xiao_SPIE09}. For ultrafast
pulse excitation, the probability of $n$ initial injected excitons
($\rho_n$) is well modeled by a Poissonian distribution
($\rho_n=\frac{\bar{n}_0^n}{n!}e^{-\bar{n}_0}$, with mean
$\bar{n}_0=A\times PI$ (/pulse/tube)).  With instantaneous
nonlinear relaxation, the PL scales with $1-\rho_0 = 1-e^{-A\times
PI}$. When the chance of injecting zero excitons is negligible,
the PL is clamped.  A fit of this function to the data in Fig 1
(solid curve on \ref{fig:PPD98D} (b)) yields an absorption
coefficient of A = 0.027 which compares with the high end of the
range of 0.003 - 0.06 reported by different groups
\cite{Krauss_NL07,Tsyboulski_NL07,Islam_PRL04,Murakami_PRL05,Randy_PRB05,Huang_PRL06,Berciaud_PRL08}.
If one compares this result only with groups who observe similar
narrow absorption linewidths, this simple model result is within a
factor of two of their estimates~\cite{Tsyboulski_NL07}.
Interestingly enough, this absorption coefficient ($A$ = 2.7\% of
photons incident on the SWCNT are absorbed) is quite similar to
the measured value for graphene: 2.3\% per layer \cite{Mak_PRL08}.

Though the above model fits the data and provides an upper bound
for absorption, it is not clear if the approximation used
(instantaneous exciton-exciton annihilation over micron-length
scales) is appropriate.  By adapting the stochastic model of
exciton relaxation, as proposed by Barzykin and
Tachiya~\cite{Barzykin_PRB05} we examine this in more detail. In
contrast to the past work that employed an initial Poissonian
distribution over different SWCNTs in an ensemble system, we model
the initial Poissonian distribution over many pulses on a single
SWCNT. Injection and relaxation from E$_{22}$ to E$_{11}$ are
assumed very fast~\cite{Manzoni_PRL05,Ma_MP06} so that initial
exciton number at E$_{11}$ is identical to the total photon number
absorbed at E$_{22}$. By numerically solving the coupled rate
equations, the average number of excitons per pulse per nanotube
as a function of time, $\bar{n}(\Gamma t)$, can be calculated,
where $\Gamma$ is the linear exciton relaxation rate. The
simulation results are compared to the experimental data as
following:
\begin{eqnarray}
PI&=& \frac{\bar{n}(0)}{A}
\label{eq:PIConversion}\\
PE&=& \left\{ \int_{0}^{\infty}\; \bar{n}(\Gamma t)d(\Gamma t)
\right\}\times\alpha\eta_{QE} \label{eq:PEConversion}
\end{eqnarray}
Note that the time variable is normalized to $\Gamma$. As a
result, $PE$ as a function of $PI$ depends only on the relative
rates of linear to nonlinear exciton relaxation
($\frac{\Gamma}{\Gamma_{NL}}$). Comparison between theory and
experiment is achieved by scaling the slope of the numerical
results in the linear regime to match the experimentally
determined PL action cross section ($A\times \alpha\eta_{QE}$);
thus the number of free parameters is reduced to two. The best fit
of simulation to experimental is determined by reduced chi-square
minimization ($\chi_r^2$) as a function of
$\frac{\Gamma}{\Gamma_{NL}}$ and $A$ (Figure \ref{fig:PPD98D}
(c)). As illustrated in Figure \ref{fig:PPD98D} (b), for four
simulated curves at $\frac{\Gamma}{\Gamma_{NL}}$ = 0, 0.01, 0.1,
0.2 that have been optimized in terms of $A$,
$\frac{\Gamma}{\Gamma_{NL}}$ = 0 and $A$ = 0.027 yield the best
fit. Note that some SWCNT exhibit a declining trend at higher pump
fluence ($PI > 250$) suggesting the presence of a higher order
nonlinear process (which is not included in the modeling), thus we
fit to below $PI = 250$. To verify this model and resulting values
for absorption and decay rate ratio, we compare the PL pump power
dependence with 150 fs pulse, 4 ps pulse, and CW excitation.
Experimentally, PL from all three scenarios overlap in the linear
regime as expected. PL from 4 ps excitation is identical (within
experimental uncertainty) to 150 fs excitation, indicating
nonlinear excitation and phase-space filling in E$_{22}$ is not
playing a role in the saturation process. Modeling exciton
dynamics with 4 ps injection requires extending the stochastic
model to explicitly include a generation term (\textit{i.e.} we no
longer assume an initial Poissonian distribution).  For the fit
parameters obtained above and a linear relaxation rate of (90
ps)$^{-1}$ (verified later), calculated PL output for 4 ps is very
similar to 150 fs excitation.  Increased PL for the same average
light intensity is predicted for longer optical pulses (durations
on the order of exciton linear lifetime); this is observed with CW
excitation which exhibits little saturation over the intensity
regime explored.

\begin{figure}[htbp]
\centering
\includegraphics[width=0.4\textwidth]{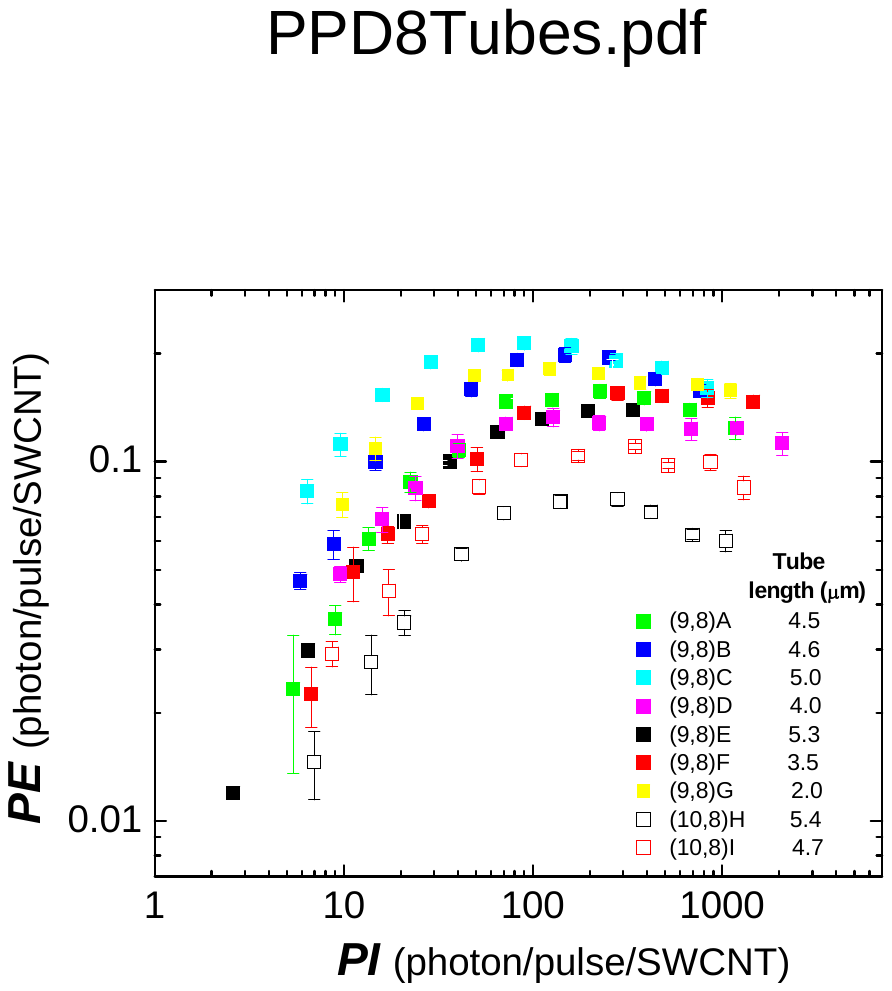}
\caption{PL emission for several (9,8) and
(10,8) SWCNTs. Error bars indicate standard error of the mean. }
\label{fig:PPDGroupingfor98and108tubes}
\end{figure}

Similar PL clamping is observed for a variety of both (9,8) and
(10,8) SWCNTs (Figure \ref{fig:PPDGroupingfor98and108tubes}).
There is some variation of PL action cross sections and maximum PL
levels, which does not correlate to length variations and is
attributed to tube imperfections still present even in these
air-suspended SWCNTs (\textit{i.e.} different $\alpha$).  Fitting
the stochastic model to the PL output from each SWCNT provides a
range of PL action cross sections: 0.002 - 0.01, absorption
coefficients: 0.02 - 0.06, and optical quantum efficiency: 7\% -
20\% (higher than the 1\% - 8\% reported by previous work
\cite{Lefebvre_NL06,Krauss_NL07,Tsyboulski_NL07,Berciaud_PRL08}.
Though the experimental uncertainty in PL action cross section is
approximately 30\% and the extraction of quantum efficiency relies
on modeling, the high values determined here are consistent with
our expectation that defects and environment effects have been
reduced in the air-suspended SWCNT.

\begin{figure}[htbp]
\centering
\includegraphics[width=0.5\textwidth]{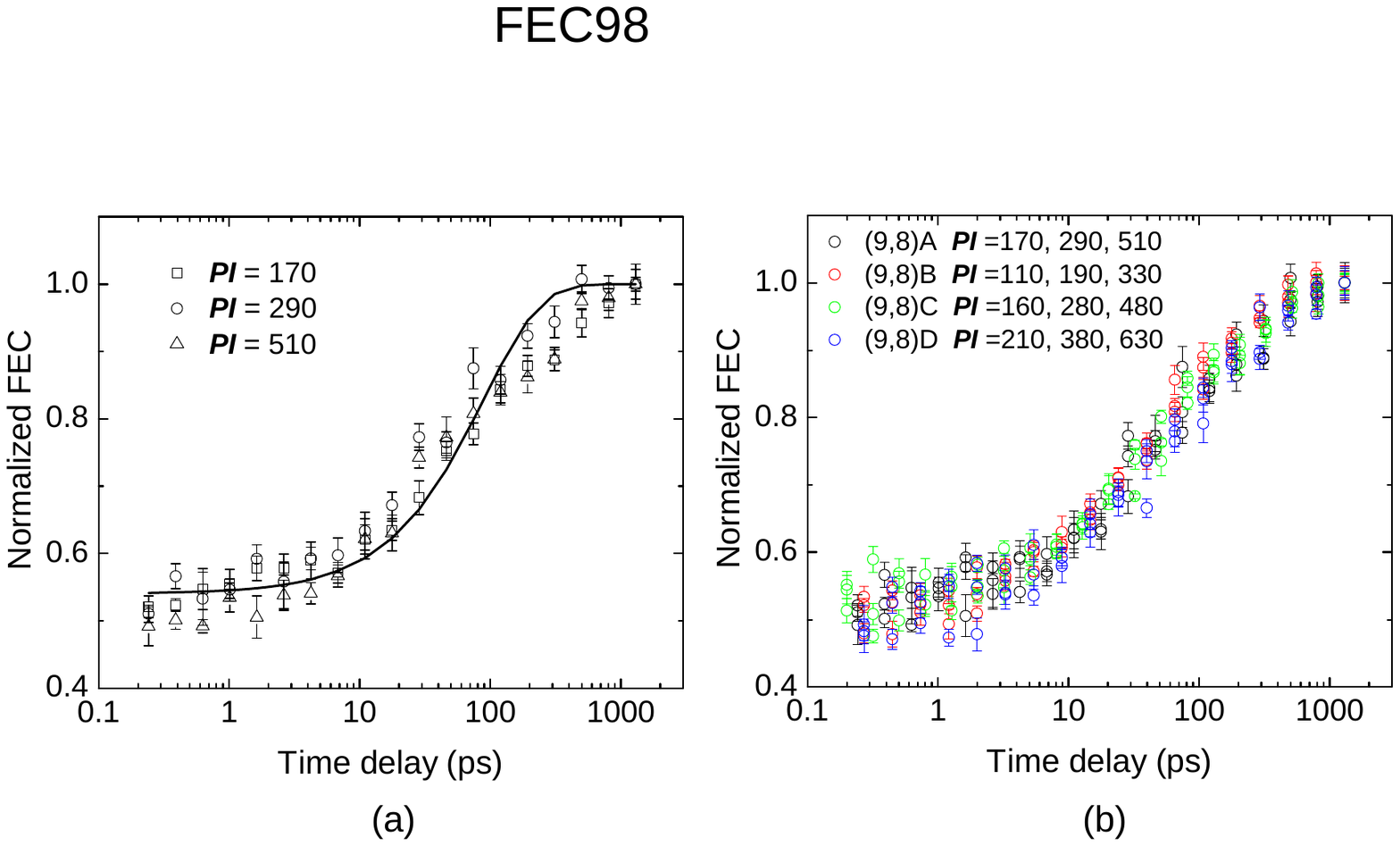}
\caption{(a) Normalized SWCNT FEC as a function of time
delay between saturating pulses for pump fluences $PI= 170,
290, 510$ . Error bars are standard deviations of the mean. Mono-exponential fit (solid curve) yields a time
constant of 90 ps. (b) Normalized FEC is very similar for
different SWCNTs and different $PI$.}
\label{fig:FEC98D}
\end{figure}

By adopting A = 0.027 for the absorption coefficient, the number
of excitons created per pulse in the SWCNT is derived (using
Eq.~\ref{eq:PIConversion}) to be around 3 at the onset of the PL
clamping. Considering the length of the nanotubes under study (2
to 5 $\mu$m), the annihilation process is much more efficient than
expected based on models developed for encapsulated
SWCNTs~\cite{Russo_PRB06,FengWang_PRB04,Murakami_PRL09}.
Application of the diffusion model to our data requires
unreasonably high absorption coefficients (orders of magnitude
higher than reported here) or dramatically longer linear radiative
lifetimes (which would allow much longer diffusion lengths).  The
analysis of PL as a function of pump fluence determines only the
ratio of $\frac{\Gamma}{\Gamma_{NL}}$. To separate linear from
nonlinear relaxation rates, we also perform time-resolved PL
relaxation experiments using femtosecond excitation correlation
spectroscopy (FEC)~\cite{Hirori_PRL06}. A single SWCNT is excited
by two equal-intensity pulses separated by variable time delay.
Since both pulses are of sufficient intensity to saturate the
SWCNT, total PL for both pulses is similar to just one pulse for
short time delays.  When time delay is increased, relaxation
between pulses can occur thus resulting in an increase in the
total PL until the point that it is double the initial value
(Figure~\ref{fig:FEC98D} (a), same (9,8)SWCNT as Fig. 1(a)).  Note
$FEC(t_d)$ is not a direct measurement of exciton relaxation. To
interpret it correctly, the FEC signal as a function of time
delay, $FEC(t_d)$, is calculated by applying the same stochastic
model used above, but for two optical pump pulses separated by
delay time $(t_d)$.  Modeling shows that for exciton decay with
two very different rates (linear and nonlinear), FEC signal is
dominated by the slow process (here the mono-exponential linear
relaxation)~\cite{Xiao_SPIE09}.  This explains the FEC signal
insensitivity to pump fluence observed in Figure~\ref{fig:FEC98D}
(a).  Phase-space filling as previously
proposed~\cite{Hirori_PRL06} is not required.  In fact, we are
more than two orders of magnitude below the expected Mott density
($> 1\times 10^6$ cm$^{-1}$)~\cite{Murakami_PRL09}.  The whole
data set is fitted to a mono-exponential function and yields a
time constant of 90 ps.

Similar results are obtained for other SWCNTs
(Figure~\ref{fig:FEC98D} (b)) with time constants ranging over
$85\pm 20$ ps. This compares to the past results of 10 - 250 ps in
encapsulated
SWCNTs~\cite{Hertel_NL05,Hagen_PRL05,Hogele_PRL08,Berciaud_PRL08}.
Since it appears that air-suspended SWCNTs have similar linear
exciton lifetimes, absorption and optical quantum efficiency as
high quality encapsulated SWCNTs, we return to their very
different nonlinear behavior. Combining the analysis results from
PL clamping and FEC, we determine the exciton-exciton annihilation
rate in our 2 - 5 $\mu$m SWCNT  ($\Gamma_{NL} > (0.01\times 85$
ps$)^{-1}$) is the same magnitude or faster than (0.8 ps)$^{-1}$
that was previously estimated for 380 nm long micelle-encapsulated
SWCNTs~\cite{FengWang_PRB04}. This suggests that with no
processing or encapsulation, exciton-exciton interactions in
air-suspended SWCNTs are affected by different microscopic
processes. Possibilities include very efficient dipole-dipole
interaction due to reduced screening, fast coherent exciton
transport due to reduced disorder~\cite{Barford_PRB06}, enhanced
coupling due to nonlocalized exciton wavefunction overlap, and
energy pooling of the excitation to only a few sites on the SWCNT
allowing annihilation at very low exciton
number~\cite{Setz_PCA05}.

In conclusion, the ability to probe an individual SWCNT suspended
in air and identify its parameters (chirality, orientation,
length) allow us to determine its intrinsic optical properties and
infer its underlying exciton dynamics. Measured PL action cross
sections are similar or higher than previously reported consistent
with high quality SWCNTs.  By investigating the dependence of PL
on pump fluence and time resolving SWCNT relaxation, we determine
that the stochastic model of exciton relaxation agrees with the
experimental results but requires an extremely fast
exciton-exciton annihilation rate. An absorption coefficient of
$A$ = 0.02 - 0.06 indicates interactions between as few as 2 - 6
excitons in a few micron long SWCNT lead to PL clamping. Extremely
efficient exciton-exciton annihilation at such low exciton density
in the quasi-1D system calls for a new consideration of the nature
of excitons optically generated in a SWCNT as well as their
interactions.

We acknowledge nanotube sample preparation and characterization by
the group including P. Finnie and J. Lefebvre at the Institute for
Microstructural Sciences, National Research Council Canada. This
work is funded by the Natural Sciences and Engineering Research
Council of Canada, the Canadian Foundation for Innovation, and the
Ministry of Research and Innovation (Ontario).


\begin{thebibliography}{99}
\bibitem{Misewich_Sci03} J. A. Misewich \textit{et al.} Science \textbf{300}, 783 (2003).
\bibitem{Freitag_NL03} M. Freitag \textit{et al.} Nano Lett. \textbf{3}, 1067 (2003).
\bibitem{Star_NL04} A. Star \textit{et al.} Nano Lett. \textbf{4}, 1587 (2004).
\bibitem{OConnell_Sci02} M. J. O'Connell \textit{et al.} Science \textbf{297}, 593 (2002).
\bibitem{Hertel_PSS06} T. Hertel \textit{et al.} Phys. Status Solidi B \textbf{243}, 3186 (2006).
\bibitem{Islam_PRL04} M. F. Islam \textit{et al.} Phys. Rev. Lett. \textbf{93}, 037404 (2004).
\bibitem{Lefebvre_PRB04} J. Lefebvre \textit{et al.} Phys. Rev. B \textbf{69}, 075403 (2004).
\bibitem{Hagen_PRL05} A. Hagen \textit{et al.} Phys. Rev. Lett. \textbf{95}, 197401 (2005).
\bibitem{Hogele_PRL08} A. H\"{o}gele \textit{et al.} Phys. Rev. Lett. \textbf{100}, 217401 (2008).
\bibitem{Weisman_Sci07} L. Cognet \textit{et al.} Science \textbf{316}, 1465 (2007).
\bibitem{Russo_PRB06} R. M. Russo \textit{et al.} Phys. Rev. B \textbf{74}, 041405 (2006).
\bibitem{Murakami_PRL09} Y. Murakami and J. Kono, Phys. Rev. Lett. \textbf{102}, 037401 (2009).
\bibitem{Luer_NP09} L. Larry \textit{et al.} Nat. Phys. \textbf{5}, 54 (2009).
\bibitem{Berciaud_PRL08} S. Berciaud \textit{et al.} Phys. Rev. Lett. \textbf{101}, 077402 (2008).
\bibitem{Lefebvre_NL06} J. Lefebvre \textit{et al.} Nano Lett. \textbf{6}, 1603 (2006).
\bibitem{Tsyboulski_NL07} D. A. Tsyboulski \textit{et al.} Nano Lett. \textbf{7}, 3080 (2007).
\bibitem{Kaminska_NT07} K. Kaminska \textit{et al.} Nanotechnology \textbf{18}, 165707 (2007).
\bibitem{calibration} $PE$ is calculated assuming isotropic SWCNT emission.  Reflection from the substrate is considered in axes calibration. Background (recorded with the SWCNT moved away from the excitation spot) has been subtracted.
\bibitem{Ma_MP06} Y.-Z. Ma \textit{et al.} Molecular Physics \textbf{104}, 1179 (2006).
\bibitem{Xiao_SPIE09} Y.-F. Xiao \textit{et al.} Proc. SPIE \textbf{7201}, 720111 (2009).
\bibitem{Murakami_PRL05} Y. Murakami \textit{et al.} Phys. Rev. Lett. \textbf{94}, 087402 (2005).
\bibitem{Krauss_NL07} L. J. Carlson \textit{et al.} Nano Lett. \textbf{7}, 3698 (2007).
\bibitem{Randy_PRB05} R. J. Ellingson \textit{et al.} Phys. Rev. B \textbf{71}, 115444 (2005).
\bibitem{Huang_PRL06} L. Huang and T. D. Krauss, Phys. Rev. Lett. \textbf{96}, 057407 (2006).
\bibitem{Mak_PRL08} K. F. Mak \textit{et al.} Phys. Rev. Lett. \textbf{101}, 196405 (2008).
\bibitem{Barzykin_PRB05} A. V. Barzykin and M. Tachiya, Phys. Rev. B \textbf{72}, 075425 (2005).
\bibitem{Manzoni_PRL05} C. Manzoni \textit{et al.} Phys. Rev. Lett. \textbf{94}, 207401 (2005).
\bibitem{FengWang_PRB04} F. Wang \textit{et al.} Phys. Rev. B \textbf{70}, 241403 (2004).
\bibitem{Hirori_PRL06} H. Hirori \textit{et al.} Phys. Rev. Lett. \textbf{97}, 257401 (2006).
\bibitem{Hertel_NL05} T. Hertel \textit{et al.} Nano Lett. \textbf{5}, 511 (2005).
\bibitem{Barford_PRB06} W. Barford and C.D.P. Duffy Phys. Rev. B \textbf{74}, 075207 (2006).
\bibitem{Setz_PCA05} P. D. Setz and R. Knochenmuss, J. Phys. Chem. A \textbf{109}, 4030 (2005).


\end{thebibliography}

\end{document}